\documentclass[10pt, a4paper, aps, eqsecnum, twocolumn, superscriptaddress]{revtex4-1}

\usepackage{amsmath,amssymb}
\usepackage{amsthm}
\usepackage{appendix}
\usepackage{comment}
\usepackage[dvipdfmx]{graphicx}
\usepackage{grffile}
\usepackage{mathrsfs}
\usepackage{bm}
\usepackage[usenames]{color}
\usepackage{ulem}
\usepackage{algorithm}
\usepackage{algorithmic}

\begin{document}
\title{\bf Use of conditional variational auto encoder to analyze ringdown gravitational waves}

\author{Takahiro S. Yamamoto}
\email{yamamoto@tap.scphys.kyoto-u.ac.jp}
\affiliation{Department of Physics, Kyoto University, Kyoto 606-8502, Japan}
\author{Takahiro Tanaka}
\email{t.tanaka@tap.scphys.kyoto-u.ac.jp}
\affiliation{Department of Physics, Kyoto University, Kyoto 606-8502, Japan}
\affiliation{Center for Gravitational Physics, Yukawa Institute for Theoretical Physics, Kyoto University, Kyoto 606-8502, Japan}

\date{\today}

\begin{abstract}
Recently, several deep learning methods are proposed for the gravitational wave data analysis.
One is conditional variational auto encoder (CVAE), proposed by Gabbard {\it et al.} \cite{Gabbard}.
We study the accuracy of a CVAE in the context of the estimation of the QNM frequency of the ringdown.
We show that the accuracy of the estimation by the CVAE is better than the matched filtering.
The areas of confidence regions are also compared and it is shown that the CVAE can return smaller confidence regions.
Also, we assess the reliability of the confidence regions estimated by the CVAE.
Our work confirms that the deep learning method has ability to compete with or overcome the matched filtering.
\end{abstract}

\maketitle


\section{Introduction}
In 2015, Laser Interferometer Gravitational-Wave Observatory (LIGO)
detected gravitational waves from a binary black hole (BBH) merger \cite{GW150914}.
In Observation run 1 and 2, ten BBH merger events were confirmed \cite{GWTC}.
Currently, advanced LIGO and advanced Virgo are operating and KAGRA will join this detector network in 2020 \cite{KAGRA}.
Besides the improvement of detectors,
the improvement of data analysis methods can contribute to accelerate the gravitational wave physics and astronomy.

Recently, the use of deep learning methods is proposed for various purposes,
e.g., the detection of gravitational waves \cite{GeorgeHuerta, SNe},
the parameter estimation \cite{Hongyu, Gabbard, Alvin}.
the noise subtraction \cite{Wei, denoiseHongyu},
and the classification of glitch noises \cite{HuertaGravitySpy}.
Our work is devoted to investigating the accuracy of the parameter estimation.
Our question is {\it how accurately deep learning methods can estimate physical parameters},
or {\it whether deep learning methods can estimate parameters more accurately than the standard method}.

In this paper, we focus on the analysis of ringdown gravitational waves.
The ringdown is the last stage of a BBH merger.
The remnant black hole is largely perturbed just after the merger and the perturbation decays as gravitational waves are emitted.
Late time perturbations of the black hole is dominated by the black hole quasi-normal modes (QNMs).
The ringdown gravitational waves can be modeled by the damped sinusoidal waveforms having the complex-valued QNM frequencies predicted by the black hole perturbation theory \cite{ReggeWheeler, Zerilli, Teukolsky}.
In general relativeity (GR), the QNM frequencies are determined by the black hole mass and spin.
Because of this property, the ringdown gravitational waves are useful for the test of GR \cite{Berti2005, LIGOtestGR}.
One way to estimate the QNM frequencies is the matched filtering using the inspiral-merger-ringdown gravitational waves \cite{GhoshA, GhoshB}. 
The posterior distribution of the binary masses and the spins is estimated and it can be converted into the mass and the spin of the remnant black hole by the fitting formula obtained from numerical relativity simulations \cite{Healy}.
This method relies on GR and the inference of these parameters is mainly governed by the inspiral part.
If the effects caused by exotic theories (e.g. modified theories of gravity, black hole mimickers)
modify the merger-ringdown part without changing the inspiral part,
bias would be introduced in the posterior in this method.
Thus, we need a method to estimate QNM frequencies using only the merger-ringdown part.

There are two possible directions of investigation:
improving the matched filtering and implementing alternative methods.
In Ref.~\cite{MDC}, comparison of various methods for the analysis of ringdown was done using test mock data.
The result shows that the deep learning method is competitive with the matched filtering.
The deep learning method used in this challenge was the one constructed for the point estimation, that is, the neural network returns only a single estimated value for each parameter that we want to estimate.
Despite of this shortcoming, deep learning methods are still expected to be a useful method complementary to the matched filtering.

Recently, the authors of Ref.~\cite{Gabbard} proposed the use of the conditional variational auto encoder (CVAE) for gravitational data analysis.
In addition to that the computational speed of the CVAE is much faster than that of the matched filtering, the CVAE can estimate the posterior probability distributions of parameters.
Although the purpose of Ref.~\cite{Gabbard} was the rapid inference,
we apply the CVAE for the off-line analysis and assess the accuracy of the inference of the CVAE.

This paper is organized as follows.
In Sec.~\ref{sec:dataset}, we present the construction of the waveforms.
In Sec.~\ref{sec:MF}, we briefly review the matched filtering.
In Sec.~\ref{sec:cvae}, the idea and the implementation of CVAE are explained.
In Sec.~\ref{sec:cnn}, we introduce convolutional neural networks (CNNs) as another competitors to the CVAE.
In Sec.~\ref{sec:results}, the results obtained by the CVAE are compared with the matched filtering and the CNN.
We focus on the accuracy of the maximum posterior estimations and the area of the confidence regions.
We also confirm that the confidence regions obtained by the CVAE have the frequentist meaning by making the P-P plot,
with evaluation of the magnitude of the error.
We summarize our results and future works in Sec.~\ref{sec:conclusion}.
Throughout this paper, we set $G=c=1$.


\section{Preparing mock templates}
\label{sec:dataset}


As explained in Introduction, the situation we consider is that only the merger-ringdown part is modified from that of GR,
and we compare deep learning methods and the matched filtering in such a situation.
For this purpose, we need to generate a test dataset by modifying only the merger-ringdown part of the waveform.
In some modified theories of gravity, gravitational waves from inspiraling BBHs can be calculated in the post-Newtonian approximation.
But consistent simulations throughout the inspiral-merger-ringdown phases have not been done so far.
In addition, it is a highly speculative assumption that only the merger-ringdown part might be modified.
Therefore, what we can do for generating modified templates is to modify the merger-ringdown phase of GR templates in a phenomenological manner.
Using the modified templates, we prepare a mock test data for comparison of the deep learning methods and the matched filtering.
These templates are used not only for preparing a test dataset, but also for training neural networks and for constructing the template bank of the matched filtering.

The precise modeling of the transition from the inspiral phase to the post-merger phase is difficult,
but we would be able to roughly assume that the gravitational waves of the merger-ringdown phase have the following properties,
\begin{itemize}
\item The amplitude after the peak monotonically decreases.
At a later time, the amplitude decays exponentially.
\item The frequency monotonically increases and converges to a certain QNM frequency at a later time.
\end{itemize}
We focus on the case where the waveforms are modified only after the time $t_p^\mathrm{GR}$, at which the amplitude of GR template reaches its peak.
Therefore, the inspiral part of the modified waveform coincides with GR one.
In this work, we focus only on $l=m=2$ mode and ignore overtones
as they are much weaker especially for nearly equal-mass binaries.
The importance of the multi-modes and overtones has been studied in Refs.~\cite{multimodeBerti, multimodeJulian, overtone}.

We denote the QNM frequencies for GR templates and for modified templates by $\omega_\mathrm{R, I}^\mathrm{GR}$ and $\omega_\mathrm{R, I}$, respectively.
The modified templates are constructed by modifying the complex-velued templates in GR, $h^\text{GR}(t)$.
First, we decompose the strain $h^\text{GR}(t)$ into the amplitude $A^\text{GR}(t)$ and the frequency $\omega^\text{GR}(t)$ as
\begin{equation}
	h^\text{GR}(t) = A^\text{GR}(t) e^{i\phi^\text{GR}(t)},\ \phi^\text{GR}(t) = \int^t dt' \omega^\text{GR}(t').
\end{equation}
From $A^\text{GR}(t)$ and $\omega^\text{GR}(t)$, the modified amplitude and frequency, $A(t)$ and $\omega(t)$, are generated.
%
%
Our modified templates are characterized by two parameters, $\delta\omega_\text{R}$ and $\delta\omega_\text{I}$.
The real and imaginary parts of the QNM frequency, $\omega_\mathrm{R}$ and $\omega_\mathrm{I}$, are specified by the fractional deviation from the GR values as
\begin{equation}
	\omega_\mathrm{R, I} = \omega_\mathrm{R, I}^\mathrm{GR} ( 1 + \delta_\mathrm{R, I}).
\end{equation}
In our work, the modifications of the frequencies are assumed to be small.
The deviations of the real part and the imaginary part of QNM frequencies are assumed to be less than 30\% and 50\%, respectively (i.e. $|\delta_\mathrm{R}|  < 0.30, |\delta_\mathrm{I}|  < 0.50$).

Modified amplitudes are constructed from two parts, before and after the peak.
After the peak, the amplitudes are modified from GR as
\begin{equation}
	A'(t) = \frac{A^\mathrm{GR}(t)}{1+e^{4 M\omega_\mathrm{I}^\mathrm{GR}x}} +\frac{A^\mathrm{RD}(t)}{1+e^{-4 M\omega_\mathrm{I}^\mathrm{GR}x}},
\end{equation}
with 
\begin{equation}
	A^{\mathrm{RD}}(t)=\frac{1.18}{1+e^{- M \omega_\mathrm{I}^{\mathrm{GR}} x}+e^{ M \omega_{\mathrm{I}} x}}.
\end{equation}
where $M$ is the total mass of the binary, $x$ is the normalized time defined as $x := (t - t^\text{GR}_p) / M$, 
and $t^\text{GR}_p$ is the time when the GR amplitude $A^\text{GR}(t)$ reaches its peak.
The time when the modified amplitude $A'(t)$ reaches its maximum is denoted by $t'_p$ and can differ from $t_p^\text{GR}$.
We connect the GR amplitude before $t_p^\mathrm{GR}$ and the modified amplitude after $t'_p$ with an appropriate normalization.
Namely, the modified amplitude $A(t)$ is obtained as
\begin{eqnarray}
	A(t) = 
	\begin{cases}
		A^\mathrm{GR} (t) & (t\leq t_p^\mathrm{GR}), \\
		\alpha A'(t+t'_p-t_p^\mathrm{GR}) & (t>t_p^\mathrm{GR}),
	\end{cases}
\end{eqnarray}
with $\alpha := A^\mathrm{GR}(t_p^\mathrm{GR}) / A'(t'_p) $.

The GW frequency $\omega(t)$ of the modified waveform is specified as
\begin{equation}
	\omega(t)=\frac{\omega^{\mathrm{GR}}(t)}{1+e^{4 M \omega_{\mathrm{I}}^{\mathrm{GR}} x}}+\frac{\omega^{\mathrm{RD}}(t)}{1+e^{- 4 M \omega_{\mathrm{I}}^{\mathrm{GR}} x}},
\end{equation}
with
\begin{equation}
	\omega^{\mathrm{RD}}(t) = \omega^\mathrm{GR}_p + (\omega_\text{R} - \omega^\text{GR}_p)\tanh (0.85 M\omega_\text{I}^\text{GR} x),
\end{equation}
and $\omega^\text{GR}_p := \omega^\text{GR}(t^\text{GR}_p)$.

Finally, we generate the gravitational wave strain, $h(t)$, by
\begin{equation}
	h(t) = A(t)e^{i\phi(t)},\ \phi(t) = \int^t dt' \omega(t').
\end{equation}
The waveform of the modified model having $\delta_\text{R} = \delta_\text{I}=0$ coincide with that of GR.

\begin{figure}[t]
\centering
\begin{minipage}{1.0\hsize}
\includegraphics[width=8cm]{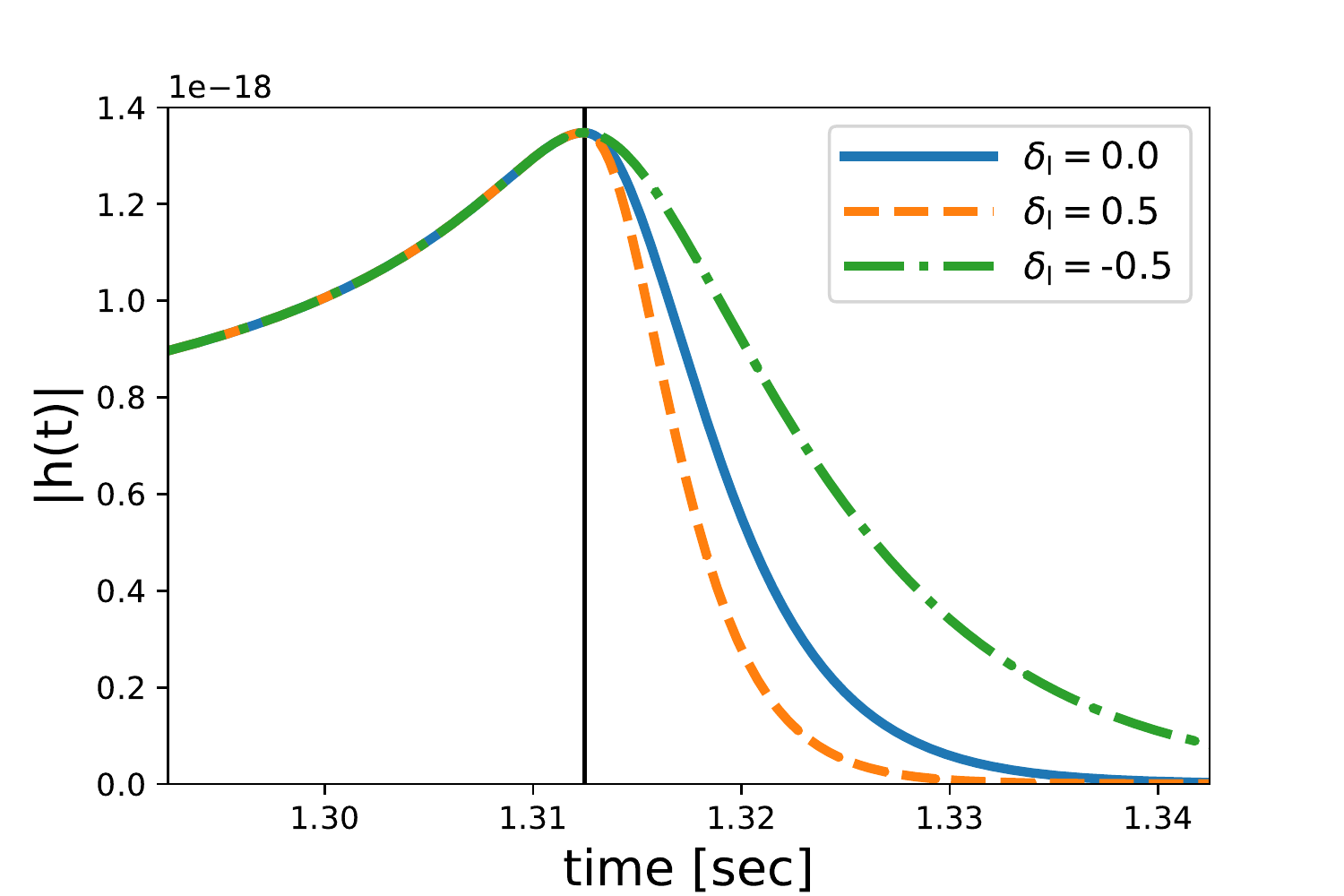}
\end{minipage}
\centering
\begin{minipage}{1.0\hsize}
\includegraphics[width=8cm]{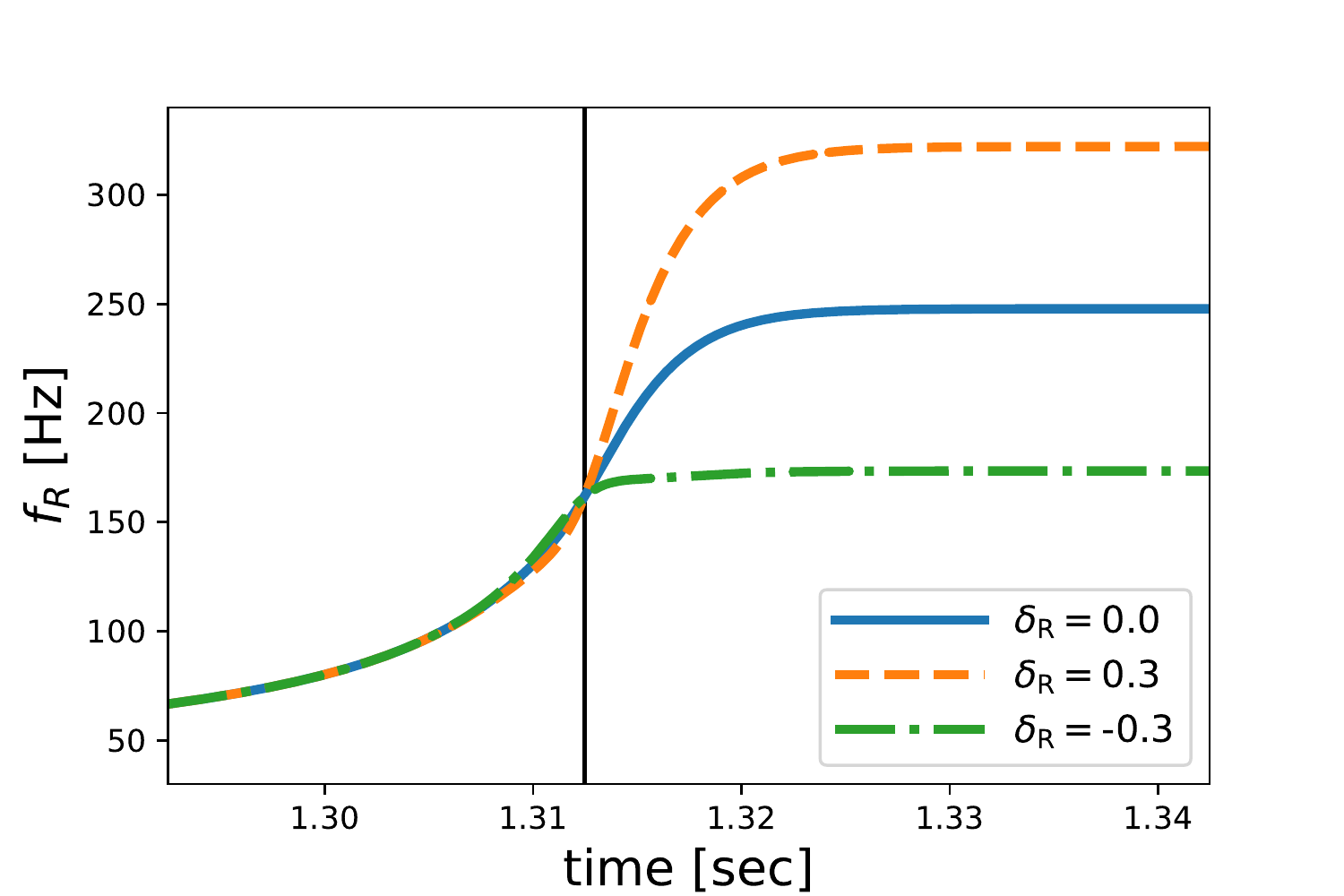}
\end{minipage}
\caption{The amplitudes ({\it top}) and the frequencies ({\it bottom}) of the modified templates having various QNM frequencies.
The frequency $f_\text{R}$ is defined as $f_\text{R} = \omega_\text{R} / 2\pi$.
When $\delta_\text{R} = \delta_\text{I}=0$, they coincide with those of GR.
The black vertical line indicates the time at which the amplitude reaches its peak.}
\label{fig:templates}
\end{figure}

As a seed for modified templates, we use the waveform SXS:0305~\cite{SXS} and the total mass is fixed to $M=72.158M_\odot$.
The GR values of QNM frequency is calculated from the fitting formula in Ref.~\cite{Berti2005}.
Examples of the modified templates are shown in Fig.~\ref{fig:templates}.

In the following analysis, the frequency $f$ is used rather than $\omega$.
They are related with each other by $\omega_\text{R,I} = 2\pi f_\text{R,I}$.
The sampling rate is 4096Hz.


\section{Matched filtering}
\label{sec:MF}
When the waveforms can be theoretically modeled and generated rapidly,
the matched filtering is a powerful method for the parameter estimation (see \cite{Creighton} as a standard textbook).
The detection statistic is the signal-to-noise ratio (SNR) and it can be calculated by the noise-weighted inner product between the observational data $s(t)$ and a template $h(t)$,
\begin{equation}
\text{SNR} = 4\text{Re} \int_{f_\text{min}}^{f_\text{max}} df\ \frac{\tilde{s}(f) \tilde{h}^\ast(f)}{S_n(f)},
\label{eq:snr}
\end{equation}
where $S_n(f)$ is the noise power spectral density,
$\tilde{s}(f)$ and $\tilde{h}(f)$ are the Fourier transforms of $s(t)$ and $h(t)$, respectively.
We use the LIGO O1 noise power spectral density,
\begin{align}
	S_n(f) = &10^{-44} \times \left( \frac{18.0}{0.1+f} \right)^4 + 0.49\times 10^{-46}  \notag \\
	&+ \left( \frac{f}{2000.0} \right)^2 \times 16.0 \times 10^{-46} [\text{strain}^2/\text{Hz}],
	\label{eq:LIGOO1}
\end{align}
given in Ref.~\cite{LIGOnoisecurve}.

We do not optimize the coalescence time in the present matched filtering analysis.
Instead, we fix it to the value of the injected templates, assuming that it can be easily guessed from the inspiral part of the gravitational wave data.
Therefore, our templates are parameterized by the deviation of the QNM frequency, $\{\delta_\text{R}, \delta_\text{I}\}$, and the initial phase, $\phi_0$.
Since the initial phase can be marginalized analytically,
the parameter search is done on the parameter space of $\{\delta_\text{R}, \delta_\text{I}\}$.
With the uniform prior, the posterior distribution of the real and imaginary parts of the QNM frequency $\{f_\text{R}, f_\text{I}\}$ can be obtained by
\begin{equation}
p(f_\text{R}, f_\text{I}|s) \propto \exp\left[ \frac{\text{SNR}^2(\delta_\text{R}, \delta_\text{I})}{2} \right].
\end{equation}

For the post-merger analysis, we set the boundaries of the integration range of frequency to $f_\text{min} = 160$Hz and $f_\text{max} = 512$Hz.
The lower cutoff frequency, $f_\text{min}$, is the frequency at which the amplitude of the template reaches the maximum.

In our work, the template bank is constructed to form a uniform grid in the $(\delta_\text{R}, \delta_\text{I})$ plane.
The parameter $\delta_\mathrm{R}$ is varied in the range $[0.7, 1.3]$ with the step of $0.006$,
while $\delta_\mathrm{I}$ in the range $[0.5, 1.5]$ with the step of $0.01$.
The template bank consists of 10,201 templates.


\section{Conditional variational auto encoder}
\label{sec:cvae}

\subsection{Idea of CVAE}
\label{subsec:ideaCVAE}

In this subsection, we explain the idea of CVAE \cite{Gabbard}.
In Bayesian inference, the existence of the true posterior $\hat{p}(y|x)$, the distribution of the physical parameters $y$ under the assumption that a signal $x$ is given, is assumed.
Here, the parameterized distributions $p_\theta(y|x)$ are used as an approximation of $\hat{p}(y|x)$.
The parameter $\theta$ depends on the input signal $x$.
The neural network is trained to estimate the relation between $x$ and $\theta$ using a training dataset,
that is, a lot of pairs of input data and the true values of the physical parameters, $\{ (x_i, y_i) \}_{i=1\dots N}$.
The Kullback-Leibler (KL) divergence,
\begin{equation}
	KL[\hat{p}(y|x) | p_\theta(y|x)] := \int dy\ \hat{p}(y|x) \log \frac{\hat{p}(y|x)}{p_\theta(y|x)},
\end{equation}
is one of the natural choices for quantifying the mismatch between two probability distributions.
Here, we consider the minimization of the expected value of the KL divergence,
\begin{equation}
	\mathbb{E}_{\hat{p}(x)}\left[ KL[\hat{p}(y|x) || p_\theta(y|x)] \right].
	\label{eq:avgKL}
\end{equation}
Because only the terms including $p_\theta(y|x)$ are essential for optimization,
the minimization of \eqref{eq:avgKL} is equivalent to the maximization of the average of the cross entropy:
\begin{align}
	&\mathbb{E}_{\hat{p}(x)}\left[ H[\hat{p}(y|x)||p_\theta(y|x)] \right] \notag \\
	&:= \int dxdy\ \hat{p}(x) \hat{p}(y|x) \log p_\theta(y|x) \notag \\
	&= \int dxdy\ \hat{p}(x, y) \log p_\theta(y|x).
	\label{eq: avgLogP}
\end{align}
This can be approximated by the sample mean,
\begin{equation}
	\mathbb{E}_{\hat{p}(x)} \left[ H[\hat{p}(y|x)||p_\theta(y|x)] \right] \simeq \frac{1}{N}\sum_{i=1}^N \log p_\theta(y_i|x_i).
	\label{eq:crossentropy}
\end{equation}

For example, Gaussian distribution can be used as $p_\theta(y|x)$.
However, it would be too simple to approximate the posterior.
In order to enhance the flexibility of the approximant, the hidden variable model is often employed.
The approximated distributions are given as a superposition of simple distributions,
\begin{equation}
	p_\theta(y|x) = \int dz\ p_{\theta_\text{D}}(y|x,z) p_{\theta_\text{E}}(z|x).
	\label{eq:hvm}
\end{equation}
The additional variables $z$, so-called \textit{hidden variables}, inherit compressed information of the data $x$.
With the hidden variable model, $\log p_\theta(y|x)$ appeared in R.H.S of Eq.~\eqref{eq:crossentropy} is bounded by the evidence lower bound (ELBO),
\begin{eqnarray}
	\log p_\theta(y|x) &\geq& \mathcal{L}_\text{ELBO} \notag \\
	&:=&\mathbb{E}_{q_\phi(z|x,y)} \left[ \log p_{\theta_\text{D}}(y|x,z) \right] \notag \\
	&&- \text{KL} \left[ q_\phi(z|x,y) | p_{\theta_\text{E}}(z|x) \right]
	\label{eq:ELBO}
\end{eqnarray}
for an arbitrary distribution $q_\phi(z|x,y)$.
The negative ELBO, $-\mathcal{L}_\text{ELBO}$, is employed as the loss function to be minimized.

A CVAE estimates the relation between the parameters of distributions and the conditioning variables.
As an example, the distribution $p_{\theta_\text{E}}(z|x)$ presents the probability of $z$ conditioned by $x$.
The neural network corresponding to $p_{\theta_\text{E}}(z|x)$ takes $x$ as an input and predicts the plausible value of $\theta_\text{E}$.
In Eq.~\eqref{eq:ELBO}, three distributions, $p_{\theta_\text{D}}$, $p_{\theta_\text{E}}$ and $q_\phi$, appear.
Therefore, we need three networks for emulating these distributions.

Further simplification of Eq.~\eqref{eq:ELBO} can be done as follows.
First, the first term of the R.H.S of Eq.~\eqref{eq:ELBO} can be approximated by the sample average,
\begin{equation}
	\mathbb{E}_{q_\phi(z|x,y)} \log p_{\theta_\text{D}}(y|x,z)
	\simeq \frac{1}{N_\text{z}}\sum_{j=1}^{N_\text{z}} \log p_{\theta_\text{D}}(y|x, z_j),
\end{equation}
where $z_j$ is the $j$-th sample of $z$ following $q_\phi(z|y,x)$.
In this work, we set $N_\text{z} = 1$.
Second, we adopt multivariate Gaussian distributions with diagonal covariance matrices as $p_{\theta_\text{D}}$, $p_{\theta_\text{E}}$ and $q_\phi$.
We denote the mean and covariance matrix of $p_{\theta_\text{E}}(z|x)$ by
\begin{subequations}
\begin{eqnarray}
	\vec{\mu}_\text{E} &=& (\mu_{\text{E}, 1}, \mu_{\text{E}, 2}, \dots, \mu_{\text{E}, D_\text{z}}), \\
	\Sigma_\text{E} &=& \text{diag}(\sigma^2_{\text{E}, 1}, \sigma^2_{\text{E}, 2}, \dots, \sigma^2_{\text{E}, D_\text{z}}),
\end{eqnarray}
those of $p_{\theta_\text{D}}(y|x,z)$ by
\begin{eqnarray}
	\vec{\mu}_\text{D} &=& (\mu_{\text{D}, 1}, \mu_{\text{D}, 2}, \dots, \mu_{\text{D}, D_\text{y}}), \\
	\Sigma_\text{D} &=& \text{diag}(\sigma^2_{\text{D}, 1}, \sigma^2_{\text{D}, 2}, \dots, \sigma^2_{\text{D}, D_\text{y}}),
\end{eqnarray}
and those of $q_\phi(z|x,y)$ by
\begin{eqnarray}
	\vec{\mu} &=& (\mu_1, \mu_2, \dots, \mu_{D_\text{z}}), \\
	\Sigma &=& \text{diag}(\sigma^2_1, \sigma^2_2, \dots, \sigma^2_{D_\text{z}}),
\end{eqnarray}
\end{subequations}
where $D_\text{z}$ and $D_\text{y}$ are the dimensions of the hidden variable $z$ and the physical parameters $y$, respectively.
Thus, the parameters $\theta_\text{E}$, $\theta_\text{D}$ and $\phi$ denoted abstractly so far are $\theta_\text{E} = \{\vec{\mu}_\text{E}, \Sigma_\text{E} \}$, $\theta_\text{D} = \{\vec{\mu}_\text{D}, \Sigma_\text{D} \}$ and $\phi = \{ \vec{\mu}, \Sigma \}$.
Then, the loss function for one training data is obtained as 
\begin{widetext}
\begin{equation}
	\text{Loss} = \frac{D_\text{y}}{2} \log 2\pi + \sum_{l=1}^{D_\text{y}} \log \sigma_{\text{D},l} + \frac{1}{2} \sum_{l=1}^{D_\text{y}} \frac{(y_l - \mu_{\text{D},l})^2}{\sigma^2_{\text{D},l}}
	+\frac{D_\text{z}}{2} - \frac{1}{2} \sum_{k=1}^{D_\text{z}} \left\{ \log \frac{\sigma^2_{\text{E},k}}{\sigma^2_{k}} + \frac{(\mu_{k} - \mu_{\text{E},k})^2}{\sigma^2_{\text{E},k}} + \frac{\sigma^2_{k}}{\sigma^2_{\text{E},k}} \right\}.
	\label{eq:loss}
\end{equation}
\end{widetext}

\begin{figure*}[t]
\centering
\includegraphics[scale=0.35]{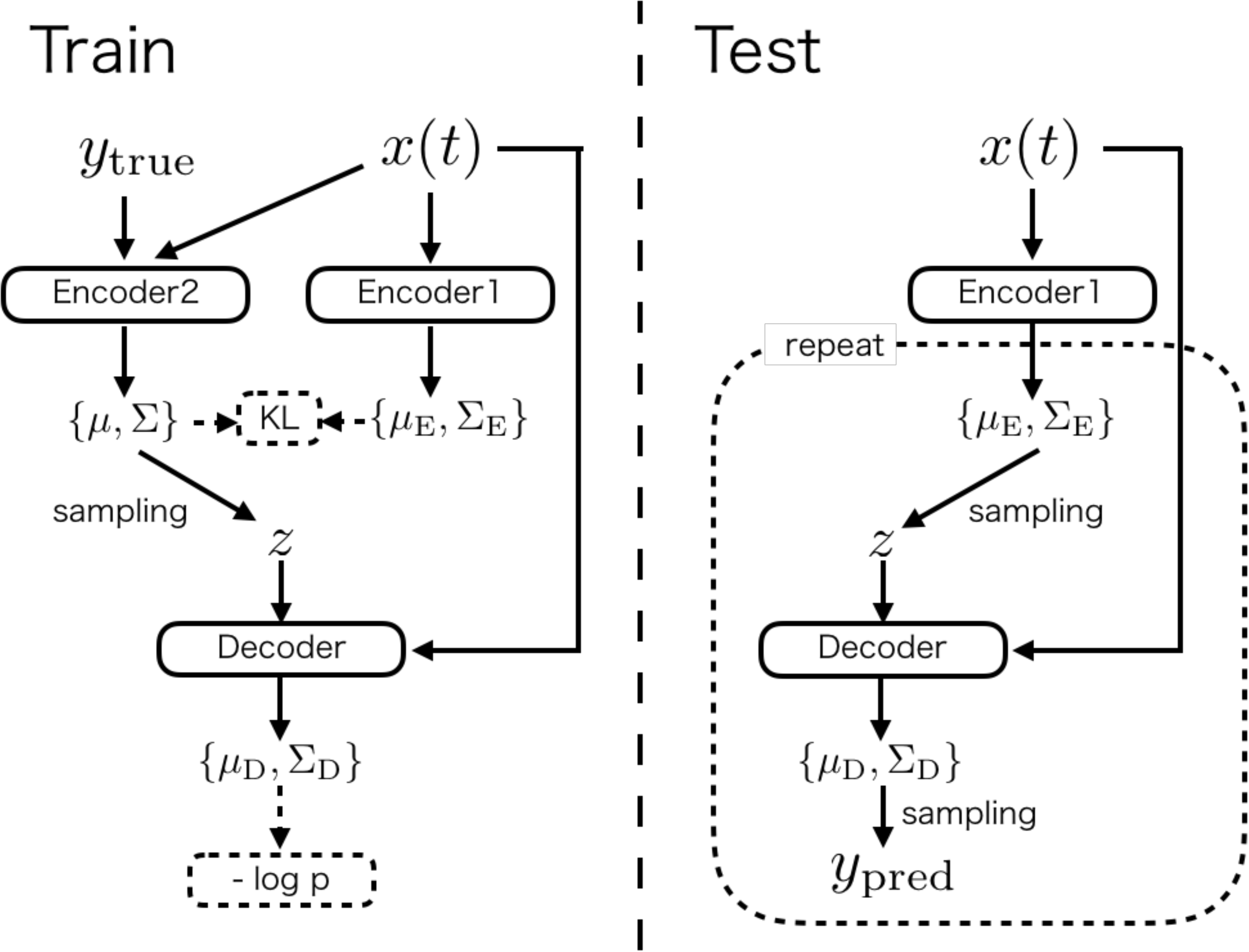}
\caption{\label{fig:cvae}
The schematic picture of the CVAE.
Encoder1, Encoder2 and Decoder represent neural networks corresponding to the probability distributions $p_{\theta_\text{E}}(z|x)$, $q_\phi(z|x,y)$ and $p_{\theta_\text{D}}(y|x,z)$, respectively.
Here, we adopt multivariate Gaussian distributions for all three distributions.
The parameters characterizing these distributions are $\theta_\text{D} = \{\mu_\text{D}, \Sigma_\text{D}\}$, $\theta_\text{E} = \{\mu_\text{E}, \Sigma_\text{E}\}$ and $\phi = \{\mu, \Sigma\}$.
At the training ({\it left}), three networks are optimized so that the loss function is minimized.
The Kulback-Leibller divergence are calculated with the output of the Encoder1 and the Encoder 2.
The output of the Decoder is used for assessing the negative log posterior term.
For test events ({\it right}), the Encoder 1 and the Decoder are employed for sampling predicted values.}
\end{figure*}

Figure~\ref{fig:cvae} shows the schematic picture of the CVAE we use in this work.
The neural networks corresponding to $p_{\theta_\text{E}}(z|x)$, $q_\phi(z|x,y)$ and $p_{\theta_\text{D}}(y|x,z)$ are called as Encoder1, Encoder2 and Decoder, respectively.
Each neural network returns the mean and the diagonal elements of the covariance matrices of each distribution.
At the training (the left figure of Fig.~\ref{fig:cvae}), all networks are simultaneously trained with the loss function \eqref{eq:loss}.
When the trained the CVAE is applied to a test data (the right figure of Fig.~\ref{fig:cvae}),
we use the networks corresponding to $p_{\theta_\text{D}}$ and $p_{\theta_\text{E}}$ for estimating a posterior.
Estimating the posterior for a test event is based on the following sampling method.
First we sample one value of $z$ from the distribution $p_{\theta_\text{E}}(z|x)$.
Next, with the sampled $z$, a sample of the parameter $y$ is obtained from $p_{\theta_\text{D}}(y|z, x)$.
Repeating these sampling processes, we finally obtain many samples of predicted values $y$ that follow the estimated posterior $p_\theta(y|x)$.

\subsection{Implementation}

In this subsection, the implementation of the CVAE that we use is described.
We use \verb|PyTorch| \cite{PyTorch} for the implementation.

\subsubsection{Structure}

As explained in the subsection \ref{subsec:ideaCVAE}, the CVAE consists of three neural networks, that is, two encoders and one decoder.
Each of them has six layers and each internal layer has 512 units.
We put a ReLU layer after each fully-connected layer except for the last layer of each neural network.
Encoder1 and Encoder2 will output the mean and the diagonal elements of the covariance matrix of the hidden variables.
We set the dimension of the hidden variables as $D_\text{z} = 16$.
The input of Decoder is the sampled variables from the multi-variate Gaussian distribution having the mean and covariance matrix estimated by the encoder.
Decoder returns the mean and the covariance matrix of the distribution $p_{\theta_D}(y|z,x)$.
The entire structure of the CVAE we use in this work is shown in Table \ref{tab:structureCVAE}.

\begin{table}[b]
\centering
\caption{\label{tab:structureCVAE}
The structure of the CAVE that we use in this work.
All layers of Encoder1, Encoder2 and Decoder are fully connected layers.
Each network consists of six fully connected layers.
The input of Encoder1 is the segment of the signal.
The inputs of Encoder2 are a segment of the signal and the injected values of $\delta_\text{R}$ and $\delta_\text{I}$.
Decoder takes the signal and the hidden variables as input.}
\begin{ruledtabular}
\begin{tabular}{cc}
Network & \# of units of respective layers\\ \hline
Encoder1 & [128, 512, 512, 512, 512, 512, 32] \\
Encoder2 & [130, 512, 512, 512, 512, 512, 32] \\
Decoder & [144, 512, 512, 512, 512, 512, 4]
\end{tabular}
\end{ruledtabular}
\end{table}

\subsubsection{Dataset for training}

For the training, we use the same templates contained in the template bank for the matched filtering.
Each template is labeled by $\{ \delta_\text{R}, \delta_\text{I}\}$.
The input signals as training data are generated as
\begin{equation}
	x(t) = A h_\text{whitened}(t) + n(t)
\end{equation}
where $h_\text{whitened}$ is a template whitened with Eq.~\eqref{eq:LIGOO1},
the noise $n(t)$ is generated from the standard normal distribution,
and the amplitude $A$ is chosen to realize a specified SNR.
To prevent overfitting to a specific noise pattern, the noise realizations are generated and the whitened templates are injected into them for each iteration.
From these simulated signals, we pick up 128 points starting from the amplitude peak,
which is used as the input data of the CVAE.

\subsubsection{Training and inference scheme}

The Adam procedure \cite{Adam} is used for the optimization algorithm.
The learning rate is set to $10^{-5}$ initially and decreased to $10^{-6}$ on the later stage of the training.
The scheduled training is employed, i.e., 
the amplitude of the signal is gradually decreased from a large initial amplitude.
The training schedule is shown in Table \ref{tab:scheduleCVAE}.
The batch size is 256.

When the trained CVAE is applied to a test data, the sampling process to estimate the distribution is repeated until $4\times10^6$ samples are collected.

\begin{table}[t]
\caption{\label{tab:scheduleCVAE}
The training schedule for the CVAE.
In the last stage of training, input signals have SNR varying from 8 to 30.
After 45000 epochs, the training is terminated when  decreasing of training loss saturates.}
\begin{ruledtabular}
\begin{tabular}{ccc}
epoch & the range of $A$ & learning rate\\ \hline
1 - 10000 & [8.0, 10.0] & $1.0\times10^{-5}$ \\
10001 - 15000 & [6.0, 10.0] & $1.0\times10^{-5}$ \\
15001 - 20000 & [4.0, 10.0] & $1.0\times10^{-5}$ \\
20001 - 25000 & [3.0, 10.0] & $1.0\times10^{-5}$ \\
25001 - 45000 & [2.0, 10.0] & $1.0\times10^{-5}$ \\
45001 -  & [2.0, 10.0] & $1.0\times10^{-6}$
\end{tabular}
\end{ruledtabular}
\end{table}

\section{Convolutional neural network}
\label{sec:cnn}

In this work, an ordinary neural network, which returns a single value for each parameter that we want to estimate,
is also implemented as one of competitors to the CVAE.
Convolutional neural networks (CNNs) are used for various research of the gravitational wave data analysis (e.g. \cite{GeorgeHuerta}).
Our CNN has three convolutional and four fully-connected layers.
Each of them, except for the last layer, is followed by a ReLU layer.
The output of the last layer is the estimated values of $\{ \delta_\text{R}, \delta_\text{I}\}$.
For respective convolutional layers, the numbers of filters are 128, 256 and 512, 
and the sizes of filters are 32, 8 and 8.
All of fully connected layers have 512 units.
We use mean square loss for the loss function.
Also for the training of the CNN, scheduled training is employed.
The training schedule is shown in Table \ref{tab:scheduleCNN}.
The CNN is also implemented by \verb|PyTorch|.
The training dataset is the same as the CVAE.

\begin{table}[t]
\caption{\label{tab:scheduleCNN}
The training schedule for the CNN.
We set the learning rate as $10^{-4}$ for the whole epoch of training.
After the 4001st epoch, the training is terminated once the decrease of training loss saturates.}
\begin{ruledtabular}
\begin{tabular}{cc}
epoch & the range of $A$ \\ \hline
1 - 1000 & [8.0, 10.0] \\
1001 - 2000 & [6.0, 10.0] \\
2001 - 3000 & [4.0, 10.0] \\
3001 - 4000 & [3.0, 10.0] \\
4001 - & [2.0, 10.0] \\
\end{tabular}
\end{ruledtabular}
\end{table}


\section{Results}
\label{sec:results}

\subsection{Dataset for comparison}
We prepare the mock test data in the same way as the training data.
The real-valued template $h_\text{inj}$ is generated from a complex-valued modified template $h = h_+ + ih_\times$ with the randomly sampled phase $\phi_0$, i.e.,
\begin{equation}
	h_\text{inj} = h_+ \cos\phi_0 + h_\times \sin\phi_0.
\end{equation}
We use the noise curve of LIGO O1 for generating the Gaussian noise (Eq.~\eqref{eq:LIGOO1}).
Three datasets with SNR of the merger-ringdown part 30.0, 15.0 and 8.0 are prepared (the definition of the merger-ringdown SNR is Eq.~\eqref{eq:snr}).
Each dataset consists of 500 simulated data whose $\delta_\text{R}$ and $\delta_\text{I}$ are randomly sampled from the region satisfying our assumptions, i.e., $|\delta_\text{R}|<0.3$ and $|\delta_\text{I}|<0.5$.

\subsection{Comparison of the point estimation}
To quantify the accuracy of the estimates, we define the following two quantities,
\begin{eqnarray}
&\overline{\Delta Q} := \dfrac{1}{N_\text{data}} {\displaystyle \sum_{i=1}^{N_\text{data}} } \left( Q_i^\text{est} - Q_i^\text{true} \right), \label{eq:DeltaQ} \\
&\sigma(Q) := \dfrac{1}{N_\text{data}} \left[ {\displaystyle \sum_{i=1}^{N_\text{data}} } \left( Q_i^\text{est} - Q_i^\text{true} \right)^2 \right]^{1/2}. \label{eq:SigmaQ}
\end{eqnarray}
Here, $Q^\text{est}$ is given by the estimated value that maximizes the posterior distribution for the matched filtering and the CVAE,
while it is given by the output value for the CNN.
The comparison of the errors is shown in Table \ref{tab:comp}.
From this table, we can conclude that
\begin{itemize}
\item For both $f_\text{R}$ and $f_\text{I}$, the means of the errors $\overline{\Delta Q}$ are much smaller than the standard deviations $\sigma(Q)$.
Therefore, the estimates of both $f_\text{R}$ and $f_\text{I}$ are not significantly biased in all methods.
\item Because the standard deviations of the CVAE are smaller than those of the matched filtering and the CNN, we can say that the CVAE estimates the QNM frequencies more accurately than the other two methods.
\end{itemize}

\begin{table}[t]
\centering
\caption{\label{tab:comp}
The comparison of the estimation errors.
The quantities $\overline{\Delta Q}$ and $\sigma(Q)$ are defined in Eqs.~\eqref{eq:DeltaQ} and \eqref{eq:SigmaQ}.
The estimation by the CVAE has no significant bias for both of $f_\text{R}$ and $f_\text{I}$ and for any values of SNR.
The matched filtering and the CNN also estimate QNM frequency with small bias for most cases.
Comparing the values of $\sigma(f_\text{R,I})$, we find that the CVAE takes the smallest values for all cases,
except for imaginary part of the dataset having SNR=8.
For this case, the CNN has a smaller value of $\sigma(f_\text{I})$ than the CVAE.
However, the CNN derives a slightly larger value of $\overline{\Delta f_\text{I}}$ than the CVAE.
This means that the estimation by the CNN is more biased.}
\begin{ruledtabular}
\begin{tabular}{llllll}
$\text{SNR}_\text{RD}$ &method & $\overline{\Delta f_\text{R}}$ [Hz] & $\sigma(f_\text{R})$ [Hz] & $\overline{\Delta f_\text{I}}$ [Hz] & $\sigma(f_\text{I})$ [Hz] \\ \hline
& MF & -0.1607 & 3.5243 & -0.1865 & 2.7237 \\
30.0 & CNN & 0.9732 & 8.2192 & -1.1812 & 3.0875 \\
& CVAE & 0.0267 & 3.1180 & -0.2528 & 2.4311 \\ \hline
& MF & -0.4015 & 7.4448 & -0.5448 & 5.4256 \\
15.0 & CNN & -0.0432 & 9.5206 & -0.6411 & 4.9630 \\
& CVAE & -0.4253 & 6.2759 & -0.2109 & 4.8657 \\ \hline
& MF & -0.1755 & 15.2181 & -1.7824 & 9.6581 \\
8.0 & CNN & 0.9783 & 14.2067 & 1.7371 & 7.7085 \\
& CVAE & -0.2350 & 12.4485 & 0.4289 & 8.9368
\end{tabular}
\end{ruledtabular}
\end{table}


\subsection{Reliability of the confidence regions}

\begin{figure}[t]
\centering
\includegraphics[width=8cm]{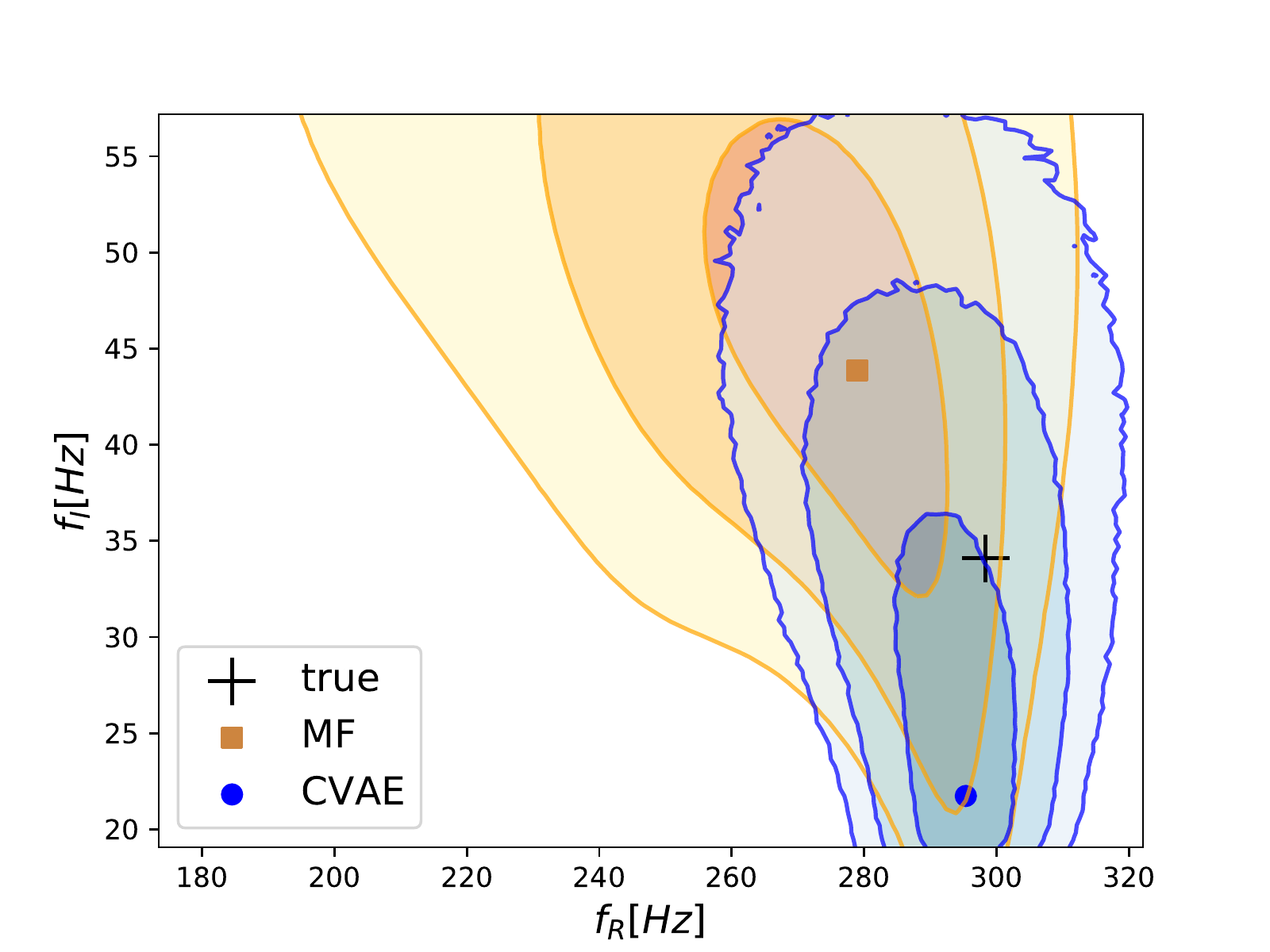}
\caption{An example of posterior estimations for a test data whose SNR is 8.0.
Blue and orange contours are confidence regions estimated by the CVAE and the matched filtering, respectively.
The contours show (50, 90, 99)\% confidence regions.
Blue circle and orange square are the predicted values of the QNM frequency obtained by the CVAE and the matched filtering, respectively.
Black cross shows the injected value of the QNM frequency.}
\label{fig:posterior}
\end{figure}

\begin{figure*}[t]
\begin{tabular}{cc}
\begin{minipage}{0.5\hsize}
\begin{center}
\includegraphics[width=8cm]{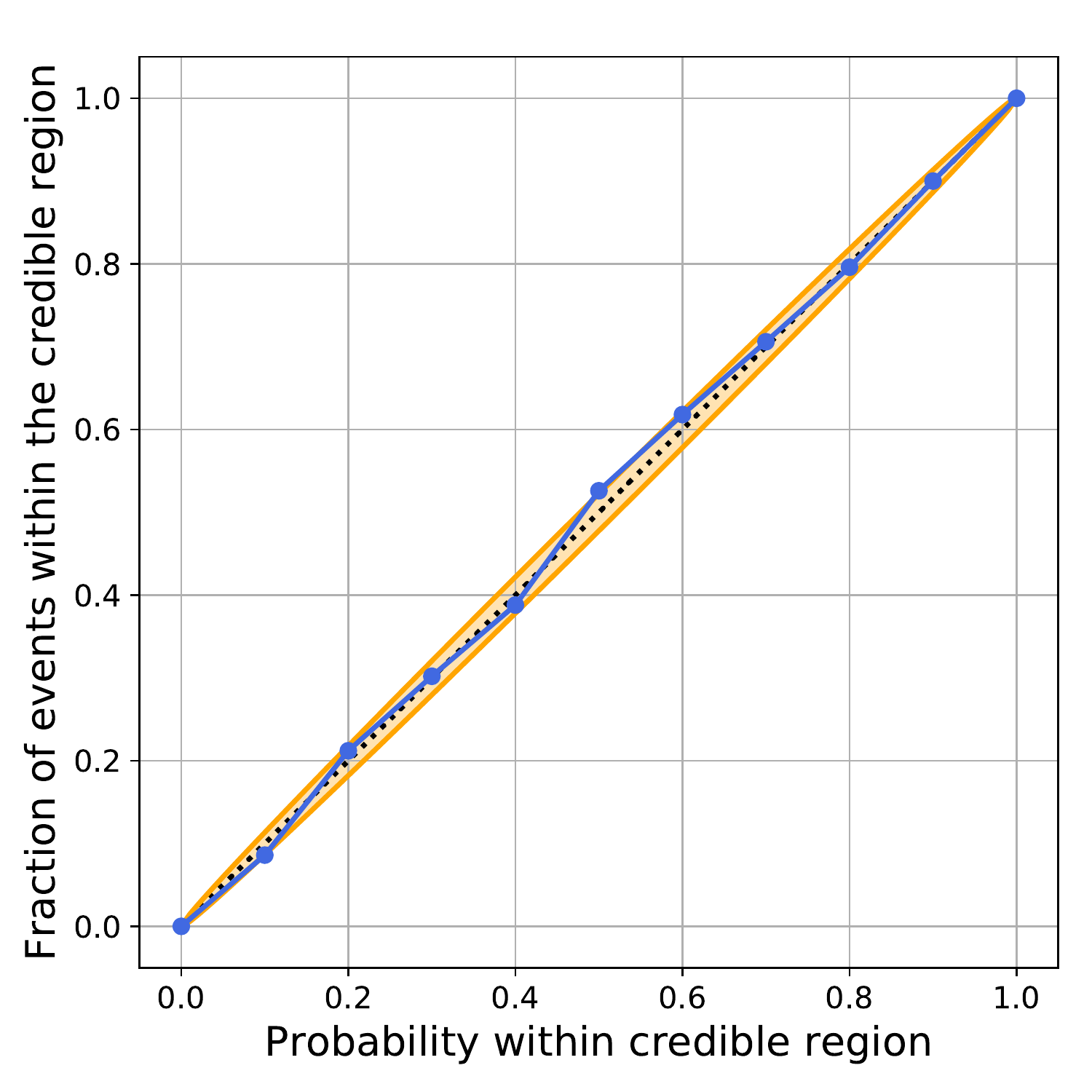}
\end{center}
\end{minipage}
\begin{minipage}{0.5\hsize}
\begin{center}
\includegraphics[width=8cm]{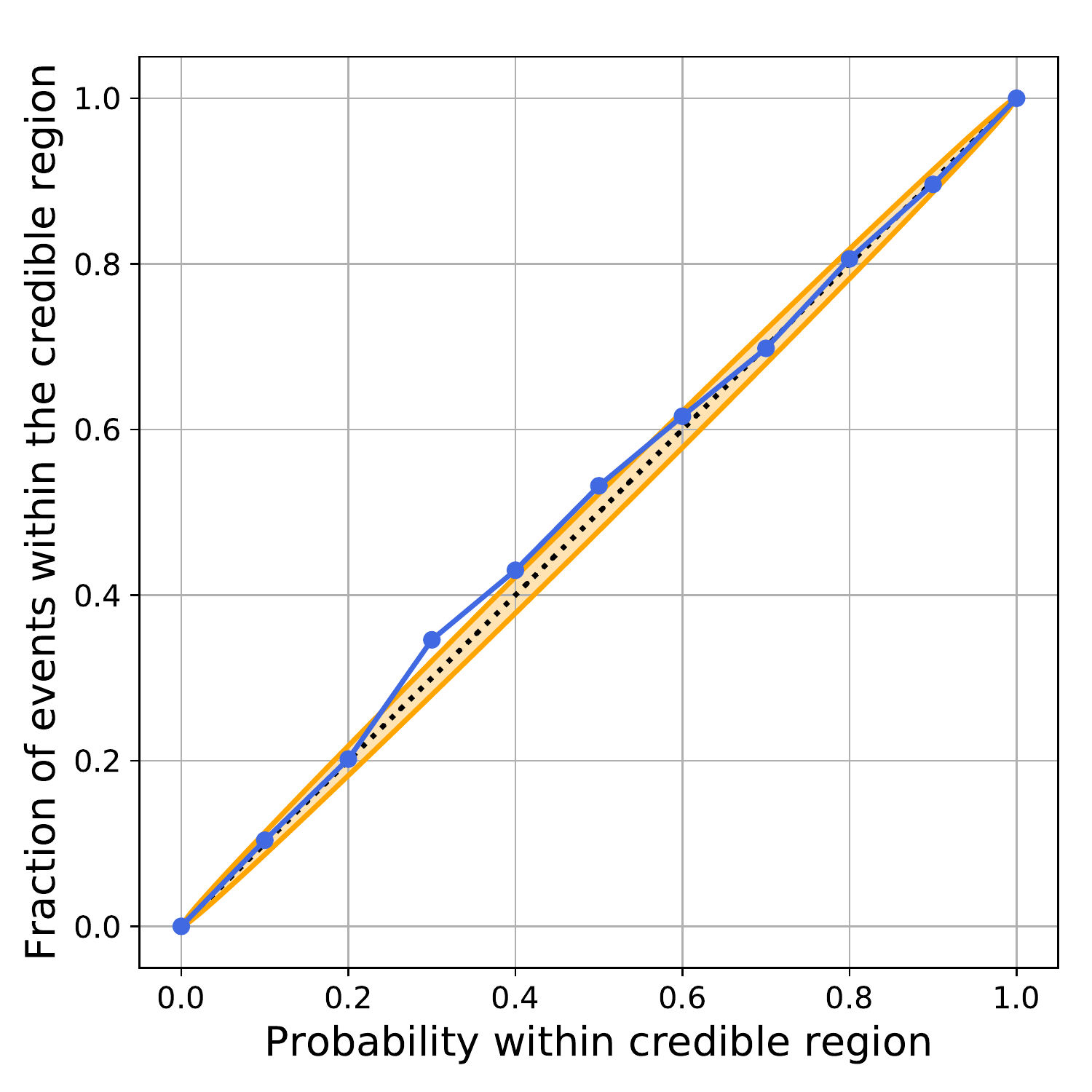}
\end{center}
\end{minipage}
\end{tabular}
\caption{\label{fig:ppplot}
The P-P plots of the matched filtering ({\it left}) and of the CVAE ({\it right}). 
The SNR of the test dataset is 8.0.
The horizontal axis shows percentages of the confidence regions.
The vertical axis shows the fraction of events whose true values are located within the confidence regions.
If estimated confidence region has the frequentist meaning, the plot (blue line) is consistent with the diagonal line (black dotted line).
The orange region is 1-$\sigma$ error of the binomial distribution.
The error estimation by the CVAE seems to be slightly biased.
A similar feature can be seen for the datasets having SNR 15.0 and 30.0.}
\end{figure*}

\begin{figure}[h]
\centering
\includegraphics[width=7cm]{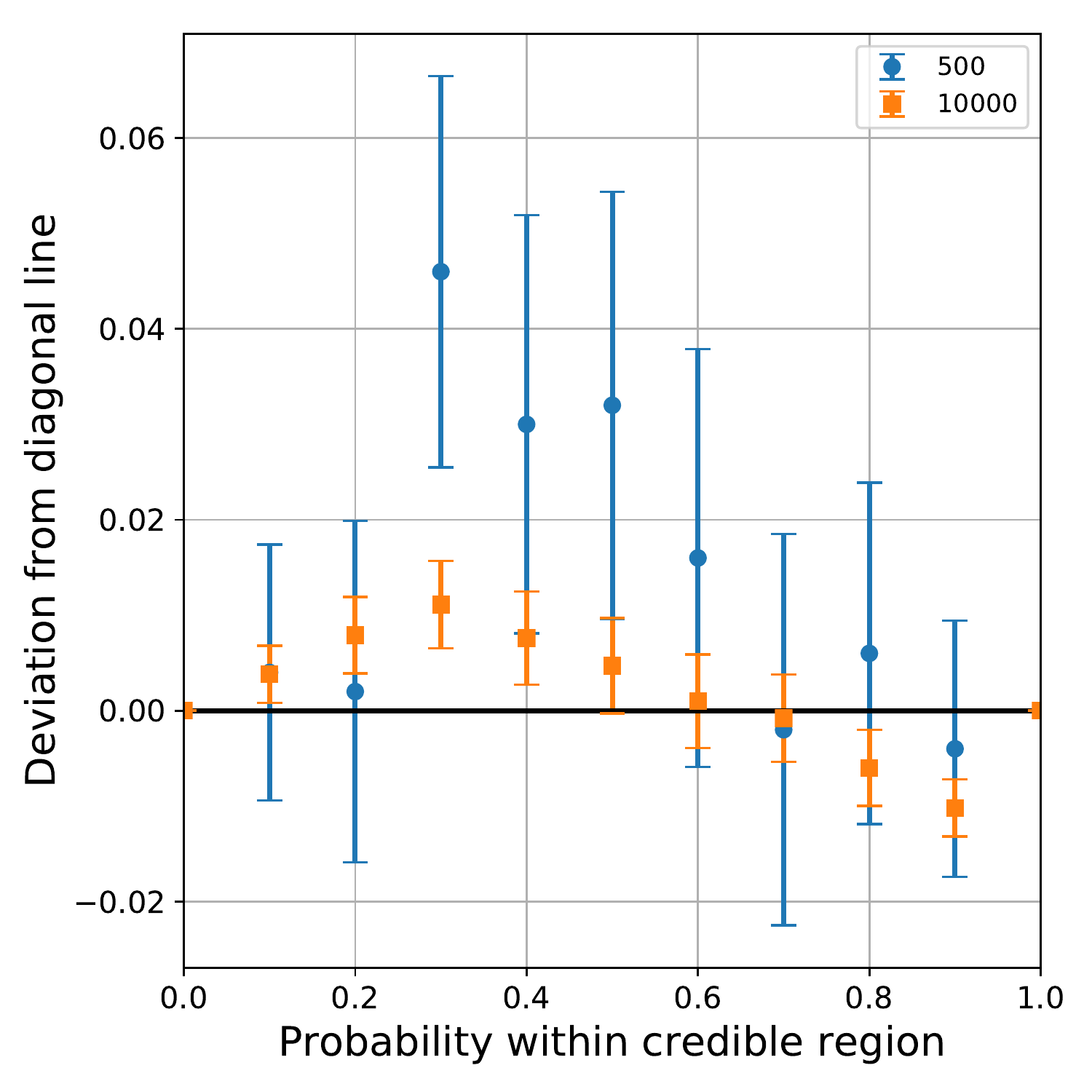}
\caption{The deviation of the P-P plot from the diagonal line.
The SNR of the dataset is 8.0.
Blue circles and orange squares are obtained with 500 and 10,000 test events, respectively.
The CVAE estimates the posterior distributions with $<2$\% systematic error.}
\label{fig:devpp}
\end{figure}

An example of the predictions of posterior distributions by the CVAE and the matched filtering is shown in Fig.~\ref{fig:posterior}.
Before comparing the posterior estimations by the CVAE and the matched filtering,
we assess the reliability of the posterior distributions estimated by the CVAE.
If the estimation of posterior distribution is reliable,
the fraction of events whose true values are located within the $x$-\% confidence region should be $x$-\%.
For visualization, a P-P plot is useful.
In a P-P plot, we take the confidence level as horizontal axis and the fraction of events as vertical axis.
If the posterior distribution is reliable, the P-P plot reduces to the diagonal line.
We show the P-P plots obtained by the CVAE and the matched filtering in Fig.~\ref{fig:ppplot}.
It is found that the error estimation by the matched filtering includes no significant bias.
On the other hand, the P-P plot for the CVAE seems to deviate from the 45$^\circ$ line only slightly.
In order to quantify the systematic error,
we generate additional 9,500 test events for each SNR.
Figure~\ref{fig:devpp} shows the deviation from 45$^\circ$ line for SNR=8.0 events.
It is found that the estimation by the CVAE contains the systematic error less than 2\%.
A similar feature can be seen for the events having SNR 15.0 and 30.0.


\subsection{Comparison of areas of confidence regions}
Taking into account the existence of bias at a few percent level,
we compare the confidence regions obtained by the CVAE and the matched filtering.
To compare them quantitatively, we define 
\begin{eqnarray}
\Delta S_i(x) = S^\text{CVAE}_i(x) - S^\text{MF}_i(x), \\
\overline{\Delta S(x)} = \frac{1}{N_\text{data}} \sum_{i=1}^{N_\text{data}} \Delta S_i(x),\label{eq:S(x)}
\end{eqnarray}
where $S^\text{CVAE/MF}_i(x)$ is the area of the $x$-\% confidence region estimated by the CVAE/the matched filtering for the $i$-th test event.
When $\Delta S_i(x)$ is negative, the constraint of the CVAE is tighter than that of the matched filtering.
The comparison of the area of the confidence region is shown in Table~\ref{tab:area}.
For all datasets, the CVAE leads to more stringent constraint than the matched filtering.

\begin{table}[t]
\centering
\caption{\label{tab:area}
The comparison of the areas of confidence regions.
The quantity $\overline{\Delta S(x)}$ is defined in Eq.~\eqref{eq:S(x)}.
For all datasets having different SNRs, the CVAE gives tighter constraint than the matched filtering.}
\begin{ruledtabular}
\begin{tabular}{llll}
$\text{SNR}_\text{RD}$ & $\overline{\Delta S(99)} [\text{Hz}^2]$ & $\overline{\Delta S(90)} [\text{Hz}^2]$ & $\overline{\Delta S(50)} [\text{Hz}^2]$ \\ \hline
30.0 & -10.8893 & -6.6020 & -2.3531 \\
15.0 & -119.521 & -64.5984 & -20.1443 \\
8.0 & -415.235 & -185.065 & -46.8837 
\end{tabular}
\end{ruledtabular}
\end{table}


\section{Conclusion}
\label{sec:conclusion}

In this paper, we investigated how accurately a CVAE can estimate the QNM frequencies using only merger-ringdown waveforms.
To do this, we generated modified waveforms by changing the merger-ringdown part of the GR template
and constructed a test dataset by injecting the waveforms into simulated Gaussian noise data.
We compared the accuracies of the CVAE and the matched filtering,
and showed the CVAE can predict the QNM frequencies with a higher accuracy than the matched filtering.
Next, we evaluated the reliability of the confidence regions estimated by the CVAE, making a P-P plot.
The estimated confidence levels have the systematic error less than 2\%.
The areas of 50\%, 90\% and 99 \% confidence regions obtained by the CVAE and the matched filtering were compared
and it was found that the CVAE can give more stringent constraint to the QNM frequencies than the matched filtering.

In this work, we only focused on the case of the Gaussian noise.
To make the deep learning method applicable to the real event analysis,
the case with the noise having non-Gaussianity need to be investigated.
The higher modes of the ringdown signal were also neglected.
The importance of the multi-mode analysis is indicated by several authors \cite{multimodeBerti, multimodeJulian}.
Application to the black hole spectroscopy is remaining for future work.

CVAE is not the only method for estimating posteriors (e.g. Bayesian neural network \cite{Hongyu}, NN with reduced order modeling \cite{Alvin}).
Comparison (or integration) with these methods would be insightful.

In this work, the merger-ringdown waveforms modified from those of GR were employed for training the CVAE.
In this sense, our method is model-dependent.
Although the post-merger templates based on the specific theory of modified gravity are not obtained so far,
the result of our work is insightful when they can be constructed.
On the other hand, exploring model independent methods is a possible direction of future work.
Even in non-GR theories, the ringdown gravitational waves would be expected to have the properties that the frequency is constant and the amplitude decays exponentially.
Neural networks would be useful to detect these features from noisy signals and estimate the QNM frequencies independently of the way of modification.


\begin{acknowledgments}
This work was supported by JSPS KAKENHI Grant Number JP17H06358 (and also JP17H06357),
{\it A01: Testing gravity theories using gravitational waves,} as a part of the innovative research area,
``Gravitational wave physics and astronomy: Genesis".
We thank the members of the A01 group for useful discussions.
Some part of calculation has been performed by using GeForce 2080Ti GPU at Nagaoka University of Technology.
\end{acknowledgments}


\end{document}